\documentclass[aps,prb,preprint,superscriptaddress,showpacs,showkeys]{revtex4}

\usepackage{graphics}
\usepackage{graphicx}
\usepackage{color}
\DeclareTextSymbol{\degre}{OT1}{23}

\begin{document}


\title{Importance of non-local electron correlations in BaNiS$_{2}$ semimetal from quantum oscillations studies}


\author{Yannick Klein}%
\email[]{yannick.klein@sorbonne-universite.fr}

\author{Michele Casula} %
\affiliation{Institut de Min\'eralogie, de Physique des Mat\'eriaux et de Cosmochimie (IMPMC), Sorbonne Universit\'e, CNRS, IRD, MNHN, 4 place Jussieu 75005 Paris, France}

\author{David Santos-Cottin}
\affiliation{LPEM - ESPCI Paris, PSL Research University; CNRS; 10 rue Vauquelin, F-75005 Paris, France}
\affiliation{Sorbonne Universit\'e; LPEM, CNRS, F-75005 Paris, France}


\author{Alain Audouard} %
\author{David Vignolles} %
\affiliation{Laboratoire National des Champs Magn\'etiques Intenses (UPR 3228 CNRS, INSA, UGA, UPS), 143 avenue de Rangueil, 31400 Toulouse}

\author{Gwendal F\`eve} %
\author{Vincent Freulon} %
\author{Bernard Pla\c cais} %
\affiliation{Laboratoire Pierre Aigrain, Ecole Normale Sup\'erieure-PSL Research University, CNRS, Sorbonne Universit\'e, Universit\'e Paris Diderot-Sorbonne Paris Cit\'e, 24 rue Lhomond, 75231 Paris Cedex 05, France}

\author{Marine Verseils} %

\author{Hancheng Yang} %

\author{Lorenzo Paulatto}%

\author{Andrea Gauzzi} %
\affiliation{Institut de Min\'eralogie, de Physique des Mat\'eriaux et de Cosmochimie (IMPMC), Sorbonne Universit\'e, CNRS, IRD, MNHN, 4 place Jussieu 75005 Paris, France}


\date{\today}

\begin{abstract}
By means of Shubnikov-de-Haas and de-Haas-van-Alphen oscillations, and \textit{ab initio} calculations, we have studied the Fermi surface of high-quality BaNiS$_2$ single crystals, with mean free path $l \sim 400 ~\text{\AA}$. The angle and temperature dependence of quantum oscillations indicates a quasi-two-dimensional Fermi surface, made of an electron-like tube centred at $\Gamma$, and of 4 hole-like cones, generated by Dirac bands, weakly dispersive in the out-of-plane direction.
\textit{Ab initio} electronic structure calculations, in the density functional theory framework, show that the inclusion of screened exchange is necessary to account for the experimental Fermi pockets. Therefore, the choice of the functional becomes crucial. A modified HSE hybrid functional with 7\% of exact exchange outperforms both GGA and GGA+U density functionals, signalling the importance of non-local screened-exchange interactions in BaNiS$_2$, and, more generally, in $3d$ compensated semimetals.
\end{abstract}

\pacs{71.20.Ps, 71.18.+y, 71.38.Cn, 72.15.Lh, 71.15.Mb}
\keywords{transition metal sulphide, semimetal, BaNiS$_{2}$, quantum oscillations, hybrid functional, ab initio methods}

\maketitle

\section{Introduction}
\label{intro}

After having been considered as a potential candidate for high-temperature superconductivity\cite{mat95,has95}, which is however absent in the doping\cite{tak94} and pressure\cite{yas99} ranges explored, the BaNiS$_2$ compound has seen a renewed attention for its fascinating spin properties, discovered only very recently\cite{Santos-Cottin16}. Indeed, angle resolved photoemission spectroscopy (ARPES) and \textit{ab initio} electronic structure calculations based on the density functional theory (DFT) have shown that the spin-orbit (SO) interactions play an important role in this material\cite{Santos-Cottin16,Sla16}. Despite the inversion symmetry of the crystal and the low $Z$ number of the nickel atoms,
an unexpected Rashba spin-orbit splitting was found in the bulk, with a Rashba spin-orbit coupling (SOC) 
as large as 26 eV \AA. This value usually corresponds to much heavier elements, or to surface physics.

\begin{figure}[!h]
   \includegraphics[width=0.95\columnwidth]{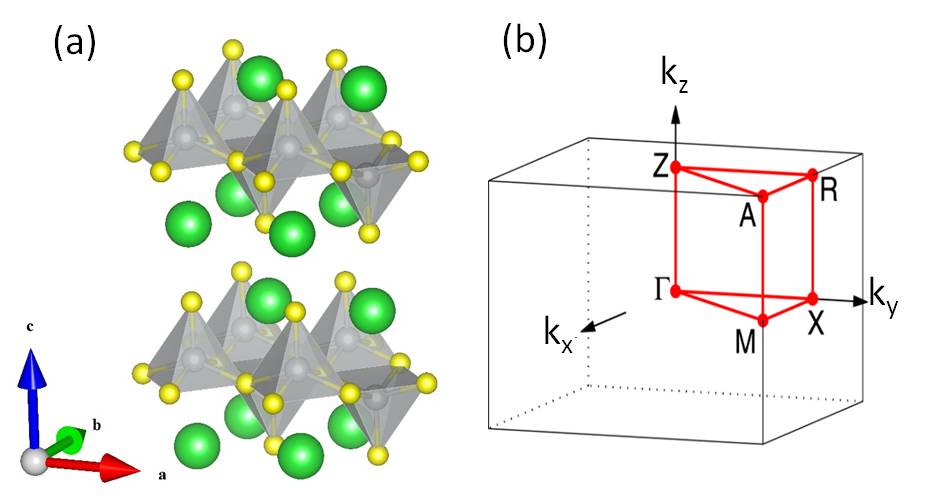}
   \caption{\label{fig1} (color online) (a) Crystallographic structure of BaNiS$_2$. Large green and small yellow spheres stand for Ba and S, respectively. Ni is represented by medium size grey spheres at the center of the pyramids (b) First Brillouin zone of BaNiS$_2$ showing high-symmetry points.} 
\end{figure}

These surprising results can be related to the peculiar symmetry of crystal structure of BaNiS$_2$, reported in Fig.~\ref{fig1}.  The transition metal site is located at the center of edge-shared square pyramids, composed of sulfur atoms, whose orientation is staggered with respect to the basal plane. The BaNiS$_2$ structure is quasi-two-dimensional, with the barium atoms acting as a charge reservoir, intercalated between the electronically active layers, 
stacked along the \textit{c} axis. The interlayer distance is approximately twice the metal-metal distance in the NiS plane. The presence of a large crystal field at the Ni sites in a locally asymmetric position enhances the SOC, and splits the electronic bands. Spin-orbit gaps and splittings of about 50 meV take place at the Fermi level, or in its proximity, paving the way to possible spintronic applications of this compound, which is a Pauli paramagnetic metal~\cite{mar95,klein16}.

In view of exploiting the unique electronic properties of BaNiS$_2$ for future spintronic applications, it is important to have a detailed account of its band structure and Fermi surface topology. Following our previous ARPES experiment of Ref. ~\onlinecite{Santos-Cottin16}, the Fermi surface can be further investigated by means of quantum oscillations (QOs). This complementary technique gives, by controlling the angle of the magnetic field, the topology of the Fermi sheets in the three-dimensional space, but without knowing their position in the first Brillouin Zone (FBZ). Except for the experimental investigations, the \textit{ab initio} description of BaNiS$_2$ can also be improved. Previous DFT calculations of the band structure of BaNiS$_2$ did not consider SOC and the Hubbard $U$ repulsion ~\cite{has95, mat95}. Nevertheless, correlation effects in BaNiS$_2$ are non negligible, owing to the presence of a transition metal element at intermediate filling. For instance, the interplay of filling, correlation, and charge transfer leads to a Fermi liquid breakdown in BaCoS$_2$, where Ni has been substituted by cobalt\cite{arxiv17}. Although our simulations within the generalized gradient approximation (GGA) supplemented by the Hubbard repulsion (GGA+U) ~\cite{Santos-Cottin16} significantly improves the description of the band structure of BaNiS$_2$ as compared to previous studies, discrepancies in the $k_z$ dispersion have been found with ARPES results.

In this paper, we bring about an improved and highly accurate description of the band structure and Fermi surface of BaNiS$_2$. For the first time, we performed Shubnikov-de Haas (SdH) and de Haas-van Alphen (dHvA) oscillation measurements on BaNi$_2$, which are direct probes for the bulk Fermi surface. The \textit{k}-space resolution in the three crystallographic directions is high enough to determine Fermi surface cross-sections with a resolution of $10^{-3}$ of the FBZ. The resolution in energy is also very high ($\Delta E \approx \mu_B B = 0.6$ meV for \textit{B} = 10 T) so that one can access to the effective mass very close to the Fermi energy. To reach these goals, we worked at low temperatures and strong magnetic fields. Moreover, we have been able to grow BaNiS$_2$ samples with a large enough scattering time ~\cite{klein16}, such that they can be investigated by means of quantum oscillations (QOs).

For the above reasons, QOs are a stringent probe to benchmark the quality of \emph{ab initio} electronic structure methods. The measured Fermi pockets volumes can be directly compared with the outcome of band structure calculations. The corresponding Fermi surface is sensitive to bandwidth renormalization, band shifts, and chemical potential. Therefore, it is a delicate quantity to compute from first principles, as it results from the fragile balance of many contributions, whose magnitude is set by the electronic correlations. On the other hand, \emph{ab initio} calculations add invaluable information to the QOs, because they allow to locate the measured Fermi pockets in the Brillouin zone, as long as a clear match can be made between measured and computed pockets.

We performed DFT band structure calculations within the GGA, GGA+U and hybrid functional approximations. We show that the local Hubbard repulsion included in a mean-field fashion in the GGA+U framework is not adequate to account for the data. Instead, we reach a qualitative agreement with the experiment only when a fraction of the exact non-local exchange is added in the functional. By appropriately tuning the Heyd-Scuseria-Ernzerhof (HSE) hybrid functional\cite{heyd2003,heyd2006}, we obtain a \emph{quantitative} agreement with the measured QOs. This corresponds to a $7\%$ of exact exchange, while the screening length $l_0$ is kept equal to the original HSE06 functional ($\omega=1/l_0=0.200 ~\textrm{\AA}^{-1}$)\cite{heyd2006,krukau2006}. The very good match between the measured and computed frequencies of the QOs allows us to provide a detailed picture of the Fermi surface topology and electronic properties of this compound. Moreover, the comparison between our three different DFT functionals against the QOs measurements reveals the importance of non-local correlation effects in this class of transition metal systems. The semimetallic character of BaNiS$_2$ is responsible for an enhanced screening of the exchange operator. However, its non-local nature cannot be neglected, because its impact on the low-energy properties and on the fermiology is strong. Indeed, the optimal value of exact exchange fraction results from electronic screening mechanisms acting on the non-local exchange operator. The stronger the metallic character, the smaller the exact exchange weight in the functional. Our 7\% value, smaller than in the original HSE functional, but still non negligible, is thus a consequence of the semimetallic character of the material.

Similar effects have been observed in the iron pnictides and selenides superconductors~\cite{Aud15,Mal14,Ter14,Wat15}, which share with BaNiS$_2$ the presence of $d$ electrons at intermediate filling, a compensated (semi)metallic character, and the $P4/nmm$ point group in the undistorted phases. Band shifts non explained by local correlations have been reported in both theory\cite{ortenzi2009} and experiments\cite{brouet2013}. Therefore, the interplay between local and non-local interaction effects is key to explain the rich physics of this class of transition metals. The present work demonstrates that QOs measurements and DFT calculations with advanced functionals can be fruitfully combined to assess and quantify in an unambiguous way the importance of these contributions.

The paper is organized as follows. In Sec.~\ref{QOs}, we detail the experimental methods (Subsec.~\ref{QOs-methods}) and report the results (Subsec.~\ref{QOs-results}) of the QOs measurements. In Sec.~\ref{elstr}, we describe the first-principles methods (Subsec.~\ref{elstr-methods}) and present the outcome of the electronic structure calculations (Subsec.~\ref{elstr-results}). We draw our conclusions in Sec.~\ref{conclusions}.


\section{Quantum oscillations measurements}
\label{QOs}

\subsection{Methods}
\label{QOs-methods}
We have grown samples by a self-flux technique similar to that reported in Ref. ~\onlinecite{sha95}, yielding high-quality platelet-like single crystals with dimensions of $\sim$ 1$\times$1$\times$0.1 mm$^3$. We selected those crystals with the best residual resistivity ratios, $\rho_{ab}(300K)/\rho_{ab}(10K) \approx 17$ for experiments in magnetic fields. We detected SdH oscillations by a four probe resistivity measurement using a lock-in detection in an Oxford Kelvinox 400 dilution refrigerator equipped with a 16-T superconducting magnet. We applied a current along the \textit{ab}-plane and the magnetic field perpendicular to it. We swept the magnetic field at a rate of 10 G/s. dHvA oscillations were deduced from magnetic torque measurement in pulsed magnetic fields of up to 58 T, using a commercial piezoresistive microcantilever ~\cite{cantilever} at the LNCMI in Toulouse. We measured variations of the piezoresistance of the cantilever with a Wheatstone bridge with an ac excitation at a frequency of 63 kHz. We varied the angle $\theta$ between the normal to the \textit{ab}-plane and the magnetic field by a one-axis rotating sample holder with a precision of +/- 1\degre.

\subsection{Results}
\label{QOs-results}
\begin{figure}[!h]
	\includegraphics[width=0.95\columnwidth]{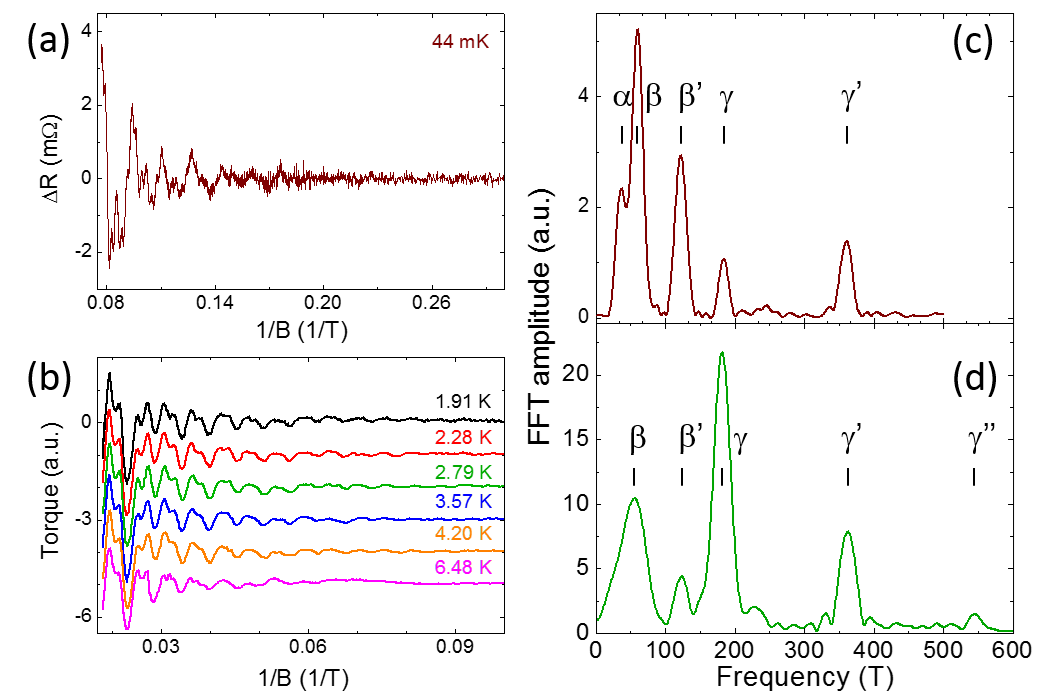}
	\caption{\label{fig2} (color online). (a) Oscillating part of the longitudinal magnetoresistance, $\Delta R$ measured at $T$ = 44 mK. (b) Oscillating part of the torque, $\Delta \tau$ at various temperatures. $\theta$ was set to 17(1)\degre. (c) and (d) display the Fourier transforms of $\Delta R$ and $\Delta \tau$ ($T = 2.28$ K) taken in the range [5 - 13 T] and [10 - 55.2 T], respectively.}
\end{figure}

Figures \ref{fig2} (a) and (b) show the oscillating parts of the longitudinal magnetoresistance and of the torque for various temperatures. Fourier analysis of the oscillatory magnetoresistance (Fig. \ref{fig2}(c)) displays peaks at frequencies $F_\alpha = 37$ T, $F_\beta = 60.5$ T, $F_{\beta'} = 122$ T, $F_\gamma = 184$ T and $F_{\gamma'} = 361$ T. These data are in agreement with Fourier analysis of the oscillatory torque (Fig.\ref{fig2}(d)) which displays peaks at frequencies $F_\beta = 60(6)$ T, $F_{\beta'} = 120(3)$ T, $F_\gamma = 182(1)$ T, $F_{\gamma'} = 363(3)$ T and $F_{\gamma''} = 549(8)$ T. 

\begin{figure}[!h]
	\includegraphics[width=0.95\columnwidth]{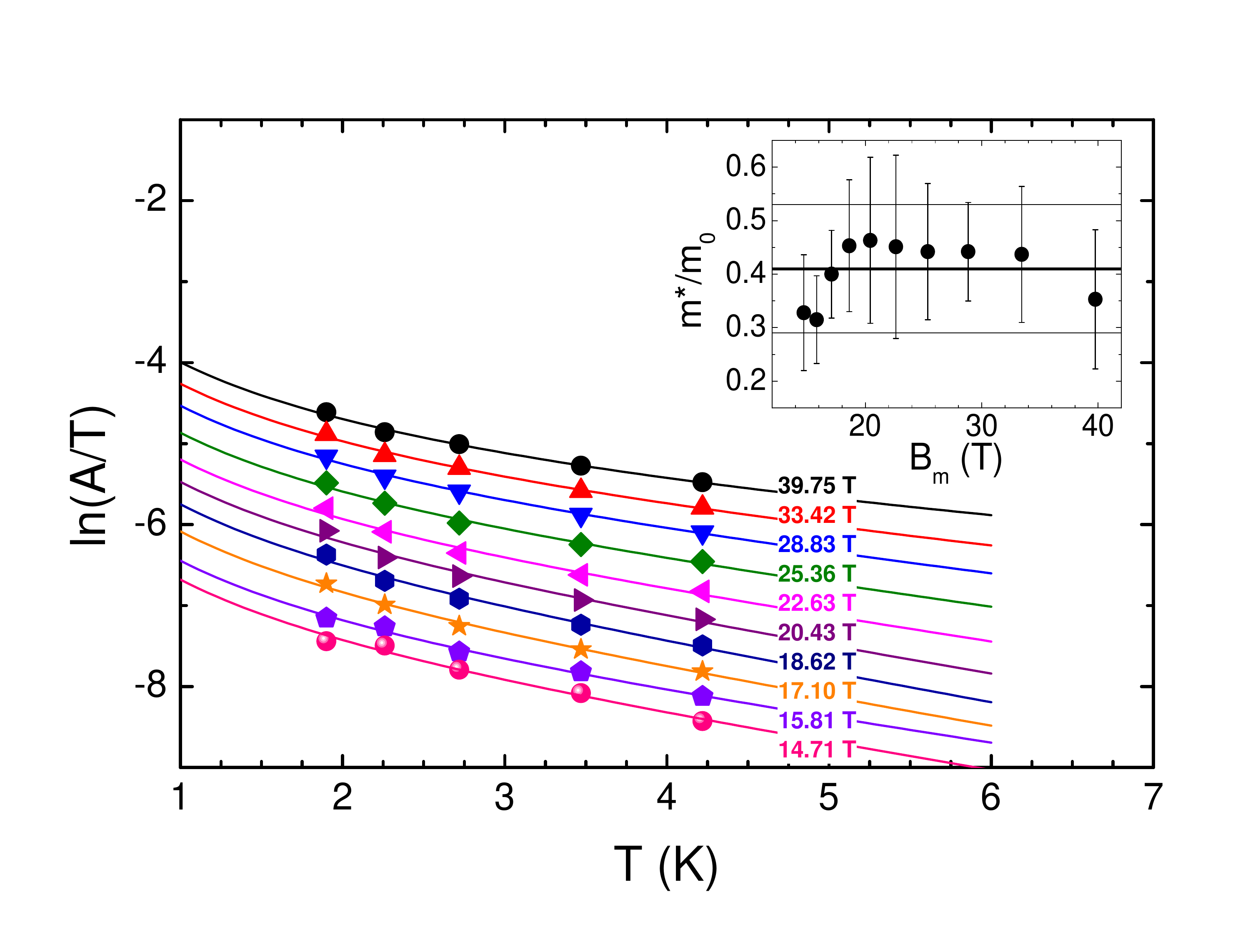}
	\caption{\label{fig3} (color online). Temperature dependence of $ln(A/T)$ where \textit{A} is the Fourier amplitude of the component with the frequency $F_{\gamma}$ determined at different mean magnetic fields, $B_m$ for $\theta = 34.5$\degre. Solid lines are the best fits of Eq. \ref{LK} to the data. Inset shows the normalized effective mass, $m^*/m_0$ issued from these fits. The mean value of $m^*/m_0$ is shown by the thick straight line.}
\end{figure}

Within the reported uncertainties, it can be considered that the frequencies $F_{\beta'}$, $F_\gamma$, $F_{\gamma'}$ and $F_{\gamma''}$ are multiple of $F_{\beta}$. Hence, it cannot be excluded that these frequencies are harmonics of either $F_{\beta}$ or $F_{\gamma}$. As discussed in the following this point can be checked through the analysis of the temperature dependence of the relevant Fourier amplitudes. Indeed, according to the Lifshitz-Kosevich (LK) formula, the amplitude of the first harmonic of a dHvA Fourier component with frequency F can be written ~\cite{shoenberg}:
\begin{equation}
\label{LK}
\Delta M \propto R_{T}R_{D}R_{S}\sin\left(\frac{2\pi F}{B}+\phi\right)
\end{equation}
where $R_{T} = \kappa T m^*/B\sinh[\kappa Tm^*/B]$, $R_{D} = \exp(-\kappa T_Dm^*/B)$ and $R_{S}$ are the thermal, Dingle and spin damping factors, respectively, with $\kappa=2\pi^2k_Bm_0/e\hbar \simeq 14.69$ T/K and $T_D = \hbar/2\pi k_B\tau$ the Dingle temperature. $m^*$ is the cyclotron effective mass in unit of the bare electron mass, $m_0$. $\tau$ is the scattering time and $\phi$ is the phase. For the $p^{th}$ harmonic, with frequency $pF$, the effective mass is $pm^*$. Temperature dependence of the Fourier amplitudes, an example of which is given in figure \ref{fig3}, are studied for three different directions of the magnetic field ($\theta =$ 17\degre, 34.5\degre, and 68\degre). The deduced effective masses are given in table \ref{table1}. 

\begin{table}[!h]
\begin{tabular}{@{}p{1cm} p{1.2cm} p{1.4cm} p{1.4cm} p{1.4cm} p{1.2cm}@{}}
\hline
\hline
$\theta$ (\degre) &	$m^*_{\alpha}$ &	$m^*_{\beta}$	& $m^*_{\gamma}$ & $m^*_{\gamma'}$  & $m^*_{\gamma''}$\\
\hline
17		& & & 0.39(12) & 0.77(8) &  \\
34.5	& $\sim 0.4$ & $\sim 0.6$ & 0.41(12) & 0.87(11) & 1.7(3) \\
68		& & 0.77(13) & 0.86(4) & & \\

\hline
\hline
\end{tabular}
\caption{Effective masses $m^*(\theta)$ in bare electron mass units for few of the Fourier components observed in figure \ref{fig2}.}
\label{table1}
\end{table}

Within the error bars, these data are compatible with the relation $m^*_{\gamma} = m^*_{\gamma'}/2 = m^*_{\gamma''}/3 $ whereas $m^*_{\gamma} \neq 3m^*_{\beta}$. It can therefore be supposed that three cyclotron orbits with frequencies $F_{\alpha}$, $F_{\beta}$ and $F_{\gamma}$ enter the Fermi surface while $F_{\beta'}$, $F_{\gamma'}$, and $F_{\gamma''}$ are harmonics. Given the uncertainty on the amplitude of $F_{\beta'}$, it was not possible to determine its effective mass.

\begin{figure}[!h]
	\includegraphics[width=0.95\columnwidth]{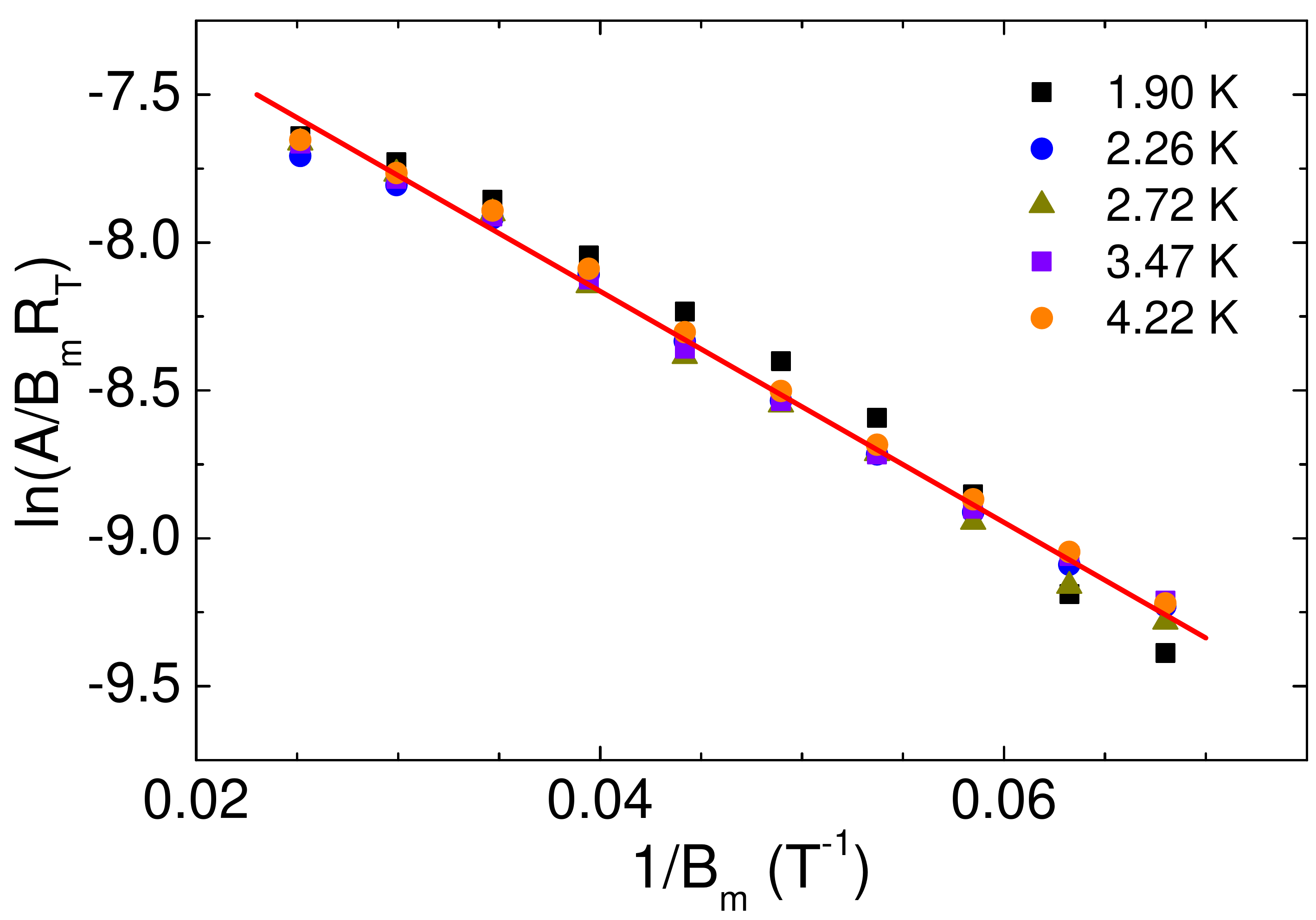}
	\caption{\label{fig4} (color online). Dingle plot for the frequency $F_{\gamma}$, at $\theta = 34.5$\degre. The solid line is the best fit of Eq. \ref{LK} with $m^*(34.5\textrm{\degre}) = 0.41 m_0$ and $T_D = 6.3$ K.}
\end{figure}

Dingle plots for the $\gamma$ orbit at $\theta = 34.5$\degre are reported in figure \ref{fig4} in which solid lines are best fits of Eq. \ref{LK} to the data with $T_D = 7$ K. Hence, the deduced scattering time is $\tau = 8.4 \times 10^{-14}$ s from which, by assuming a circular orbit with $v_F=\hbar k_F/m^*$, a mean free path, $l \sim 400 ~\text{\AA}$ can be deduced. The latter value is close to the longest mean free paths measured in high-$T_{C}$ cuprates ~\cite{seb14} and iron-based pnictides ~\cite{ana09}.

\indent  

\begin{figure}[!h]
	\includegraphics[width=0.95\columnwidth]{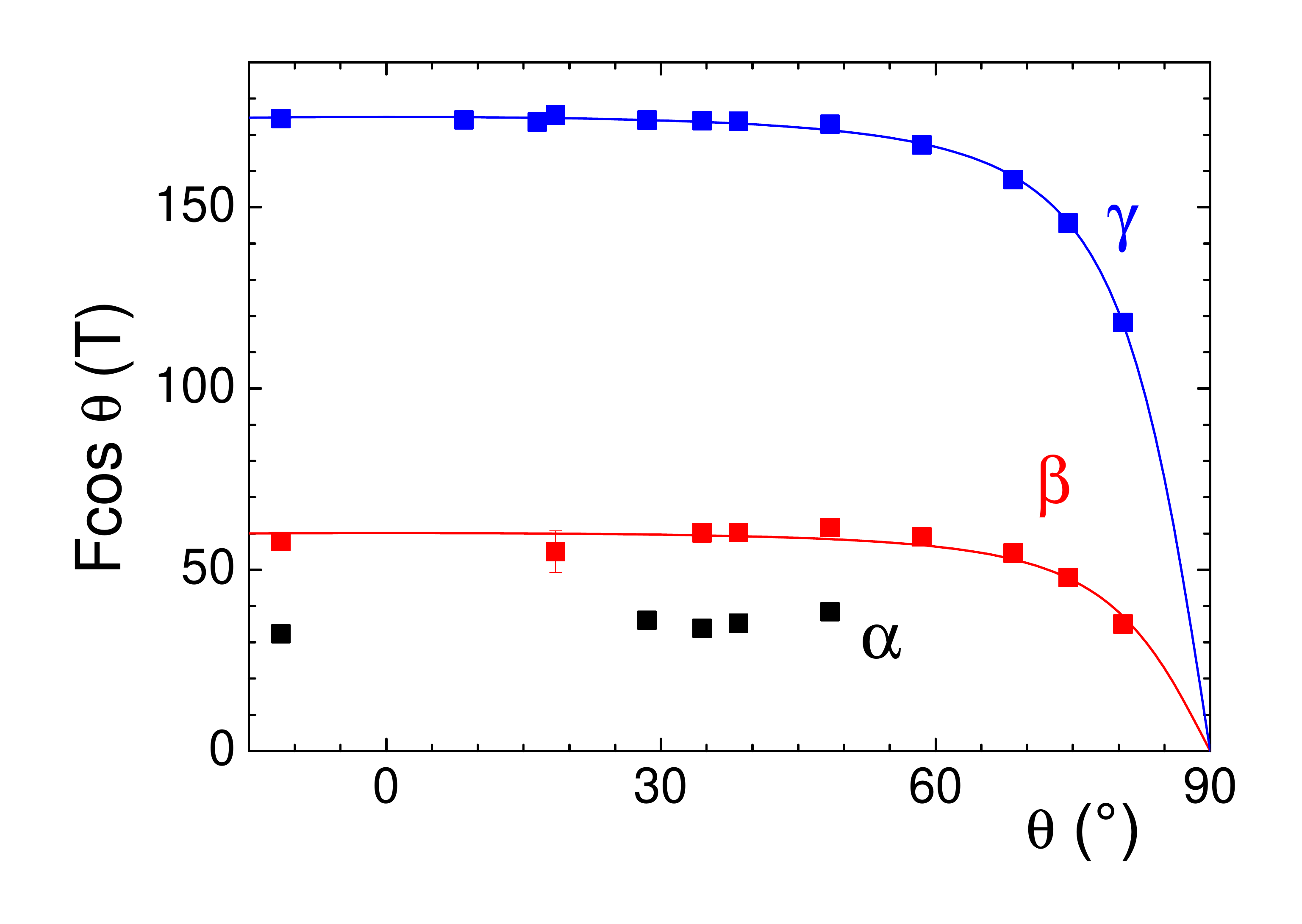}
	\caption{\label{fig5} (color online). Observed dHvA frequencies plotted as $F\cos(\theta)$ \textit{vs} $\theta$. Solid lines are best fit of Eq. \ref{ellipsoid_F} to the data relevant to $\beta$ and $\gamma$ orbits.
	} 
\end{figure}

The dimensionality of the Fermi surface can be studied by analyzing the angle dependence of the frequencies. A strictly 2D Fermi surface would result in a $\theta$-independent value of $F\cos(\theta)$. As shown by figure \ref{fig5}, this is not the case for ${\beta}$ and ${\gamma}$, since $F\cos(\theta)$ decreases as  $\theta$ increases. Figure \ref{fig5} also indicates that ${\beta}$ and ${\gamma}$ orbits belong to two different pockets because otherwise the two curves would cross at a given Yamaji angle ~\cite{yamaji}. One may also wonder if the ${\alpha}$ orbit belongs to the same Fermi surface as the $\beta$ or $\gamma$ orbits. In fact, the data in Table \ref{table1} show similar effective masses for the three orbits. As a consequence, it is possible that two of these orbits are extreme orbits of the same sheet. This point will be discussed later based on the angle dependence of the frequencies and on band calculations. It can be noticed that the angle dependence of $F_{\beta}$ and $F_{\gamma}$ is nicely accounted for by the expression :
\begin{equation}
\label{ellipsoid_F}
F(\theta)\cos(\theta) = \frac{F(0)}{\sqrt{1+\xi^2 \tan^2(\theta)}}
\end{equation}
obtained for Fermi surface pockets with a bi-axial ellipsoidal shape derived from standard parabolic bands, where $\xi = k_F^{xy}/k_F^{z}$ is the ratio of the major to minor semiaxes. Here $k_F^{z} > k_F^{xy}$ because $F\cos(\theta)$ decreases with increasing $\theta$. Assuming a circular cross section in the $k_{x}k_{y}$- plane, $k_F^{xy}$ is related to $F(0)$ through the Onsager relation : $k_F^{xy} = \sqrt{\frac{2e}{\hbar}F(0)}$. At small enough $\theta$ values, in particular for $\theta =$ 17\degre and 34\degre, both frequencies and effective masses are reasonably well approximated by a 2D Fermi surface, i.e. $F(0)=F(\theta)\cos(\theta)$ and $m^*(0)=m^*(\theta)\cos(\theta)$. While for $\theta =$ 68\degre it is necessary to consider the ellipsoidal shape: 
\begin{equation}
\label{ellipsoid_m}
m^*(\theta)\cos(\theta) = \frac{m^*(0)}{\sqrt{1+\xi^2 \tan^2(\theta)}}
\end{equation}
introducing $\xi$ values found in Eq. \ref{ellipsoid_F}. $F(0)$, deduced cross section areas, $m^*(0)$ and $\xi$ values are given in table \ref{table2}. All the orbits have very small area which is in agreement with the low carrier densities of BaNiS$_{2}$ given by band structure calculations ~\cite{mat95} and magnetotransport measurements ~\cite{klein16}.  
$\xi$ values of 0.24(2) and 0.179(4) correspond to $k_F^{z}/c^*$ ratios of 0.51(6) and 1.15(4) for the $\beta$ and $\gamma$ pockets, respectively. Therefore the $\beta$ pocket can be modeled by an elongated ellipsoid. On the other hand, $k_F^{z}/c^* > 1$ obtained for $\gamma$ is not consistent with an ellipsoid. It probably means that the ellipsoid model describes the dispersion only up to $\theta \approx 80$\degre but not more. Instead, it can be assumed that $\gamma$ is the belly orbit of a two-dimensional warped tube running along \textit{c*}. Within this picture, ${\alpha}$ could be the neck of this warped tube.

\begin{table}[!h]
\begin{tabular}{@{}p{1.5cm} p{1.5cm} p{1.8cm} p{1.5cm} p{1.2cm}@{}}
\hline
\hline
Orbit &	$F(0)$ (T) &	\% of FBZ	& $\xi$ & m$^*$(0) \\
\hline
$\alpha$	&	36(7)  &  0.17(4) &	& $\sim 0.3$ \\
$\beta$		&	63(3)  &  0.30(2) & 0.24(2) & 0.34(6) \\	
$\gamma$	&	176(1) &  0.839(5)& 0.179(4) & 0.35(8)\\	
\hline
\hline
\end{tabular}
\caption{Parameters relevant to the three detected orbits projected in the $k_{x}k_{y}$-plane. The cross section areas are given in percentage of the FBZ area. The ratio of the major to minor semiaxes ($\xi$) are deduced from the data in figure \ref{fig5} assuming ellipsoidal Fermi surfaces (see Eq. \ref{ellipsoid_F}). Effective masses are in bare electron mass units.} 
\label{table2}
\end{table}

In summary, the quantum oscillations data of BaNiS$_{2}$ are consistent with a Fermi surface composed of two pockets. The larger one, named $S_1$ is a 
constricted tube with neck and belly orbits $\alpha$ and $\gamma$, respectively. The angle dependence of the corresponding frequencies is in agreement with a sinusoidal dispersion along $k_z$, which at small enough angle between the field direction and the normal to the conducting plane can be approximated by an ellipsoid. A better resolution of $F_\alpha$ and a more detailed exploration of quantum oscillations as a function of $\theta$ would be necessary to investigate the existence of a Yamaji angle. The total volume of $S_1$ assuming such an energy dispersion is $V_1 = 6.4(6) \times 10^{-3} ~\text{\AA}^{-3}$. The smaller pocket, denoted as $S_2$ is an elongated ellipsoid which is responsible of a single frequency $F_\beta$ in quantum oscillations and which covers half of the FBZ in the $k_z$ direction for a total volume $V_2 = 1.5(4) \times 10^{-3}~\text{\AA}^{-3}$. 
\indent 

In order to establish a model of Fermi surface, we cross-check this analysis with previously published results of photoemission spectroscopy and magnetotransport measurements ~\cite{Santos-Cottin16, klein16}. According to the Hall effect data ~\cite{klein16}, BaNiS$_2$ is a semimetal very close to compensation, with a charge imbalance of only 1\% in favor of electrons.  For the compensation to be fulfilled, $S_1$ and $S_2$ must have different electronic characters and, since $V_1 \approx 4V_2$ and owing to the tetragonal symmetry of BaNiS$_2$, the former must be located at the center of the FBZ while the latter has to be repeated four times. Within this picture, the total number of holes and electrons is of the order of $\sim 5 \times 10^{19}$ cm$^{-3}$, in agreement with magnetotransport data. These statements are also in agreement with ARPES measurements ~\cite{Santos-Cottin16} which, for a given $k_z$ value, reveal two pockets: an electron-like one at the center of the FBZ and a hole-like one along $\Gamma$-M at mid-distance between the two high-symmetry end points. 

\section{\textit{ab initio} electronic structure calculations}
\label{elstr}

\subsection{Methods}
\label{elstr-methods}

In order to identify the Fermi pockets detected by QOs, we are going to use three different density functionals,
namely the GGA with the Perdew-Burke-Ernzerhof (PBE) functional\cite{perdew1996}, the PBE supplemented by local Hubbard interactions (GGA+U)\cite{anisimov1991,liechtenstein1995}, and a modified hybrid Heyd-Scuseria-Ernzerhof (HSE) functional\cite{heyd2003,heyd2006} with an optimized $7\%$ of exact exchange. 
Nowadays, the hybrid functionals, introduced for molecular systems\cite{krukau2006}, are getting more and more popular in solid state physics as well\cite{paier2006,marsman2008,franchini2014}, because they correct for some deficiencies of the regular density functionals, thanks to a reduced self-interaction error, and the restoration of the derivative discontinuity. The hybrid functionals are non-local and orbital dependent.
In all three cases, we included nickel spin-orbit (SO) interactions, whose importance has been recently demonstrated in this system\cite{Santos-Cottin16,Sla16}. Indeed, despite the quite low $Z$ number of the nickel atoms, the peculiar crystal structure of BaNiS$_2$ enhances the SOC, with sizable splittings showing up in certain regions of the Brillouin zone, and for some bands crossing the Fermi level\cite{Santos-Cottin16}. Therefore, we expect that the shape of some Fermi pockets will be affected by SOC.

All \emph{ab initio} DFT calculations have been carried out by using the Quantum ESPRESSO package\cite{QE-2009,giannozzi2017}. The geometry of the cell and the internal coordinates are taken from the experimental values reported previously ~\cite{Grey70}. Ni, Ba and S atoms are described by norm-conserving pseudopotentials. The Ni pseudopotential is fully relativistic, with 10 valence electrons (4s$^2$ 3d$^8$) and non-linear core corrections. For Ba, the semi-core states have been explicitly included in the calculations. The S pseudopotential is constructed with the 3s$^2$ 3p$^4$ in-valence configuration. The $k$-point sampling convergence is reached for a 8$\times$8$\times$8 electron-momentum grid and a Methfessel-Paxton smearing of 0.01 Ry. The plane-waves cutoff has been set to 120 Ry for the wave function in the GGA and GGA+U calculations. In the HSE calculations, we lowered this cutoff to 60 Ry in order to reduce the computational burden of the non-local exchange potential evaluation. We checked that this lower cutoff does not affect the results within a target accuracy of $0.01$ eV in the band structure.

For the GGA+U calculations performed in the fully rotational invariant framework\cite{liechtenstein1995}, we took the same Hubbard parameters as in Ref.~\onlinecite{Santos-Cottin16}, namely $U$=3 eV, $ J$=0.95 eV, and with the F4/F2 Slater integral ratio equal to its atomic value. This is in agreement with experimental values\cite{krishnakumar2001} and recent cRPA estimates\cite{arxiv17}.

For the HSE functional, we used the latest fast implementation of the exact Fock exchange energy\cite{giannozzi2017}, based on the adaptively compressed exchange (ACE) scheme\cite{lin2016}. The significant CPU time reduction, allowed by the ACE implementation, makes hybrid functional calculations of BaNiS$_2$ possible in a plane-waves framework, which guarantees a systematic and controlled basis-set convergence. The $q$-integration of the non-local Fock operator has been performed in a downsampled grid made of 8$\times$8$\times$2 $q$-points. To minimize the number of non-equivalent momenta in the $k+q$ grid, we found that shifting the $k$-grid by half-a-grid step in the $z$ direction further speeds up our calculation. The loss of precision caused by the $q$-downsampling is critical for metals. However, given the quasi-two-dimensional nature of BaNiS$_2$, reducing the $q$ grid by a factor of 4 in the out-of-plane direction does not deteriorate the results, which still fulfill our target accuracy, while leading to a significant gain in CPU time.

As a final step, we obtained an accurate determination of the band structure, chemical potential, Fermi surface, and Fermi pocket cross-sections by the Wannier interpolation of the \emph{ab initio} bands, performed with the Wannier90 code\cite{mostofi2014} for all functionals taken into account. The tight-binding model comprises 22 bands, originating from all Ni $d$ and S $p$ states in the BaNiS$_2$ unit cell.

\subsection{Results}
\label{elstr-results}

\begin{figure}[!h]
\centering 
\begin{tabular}{cc}  
	\includegraphics[width=0.48\columnwidth]{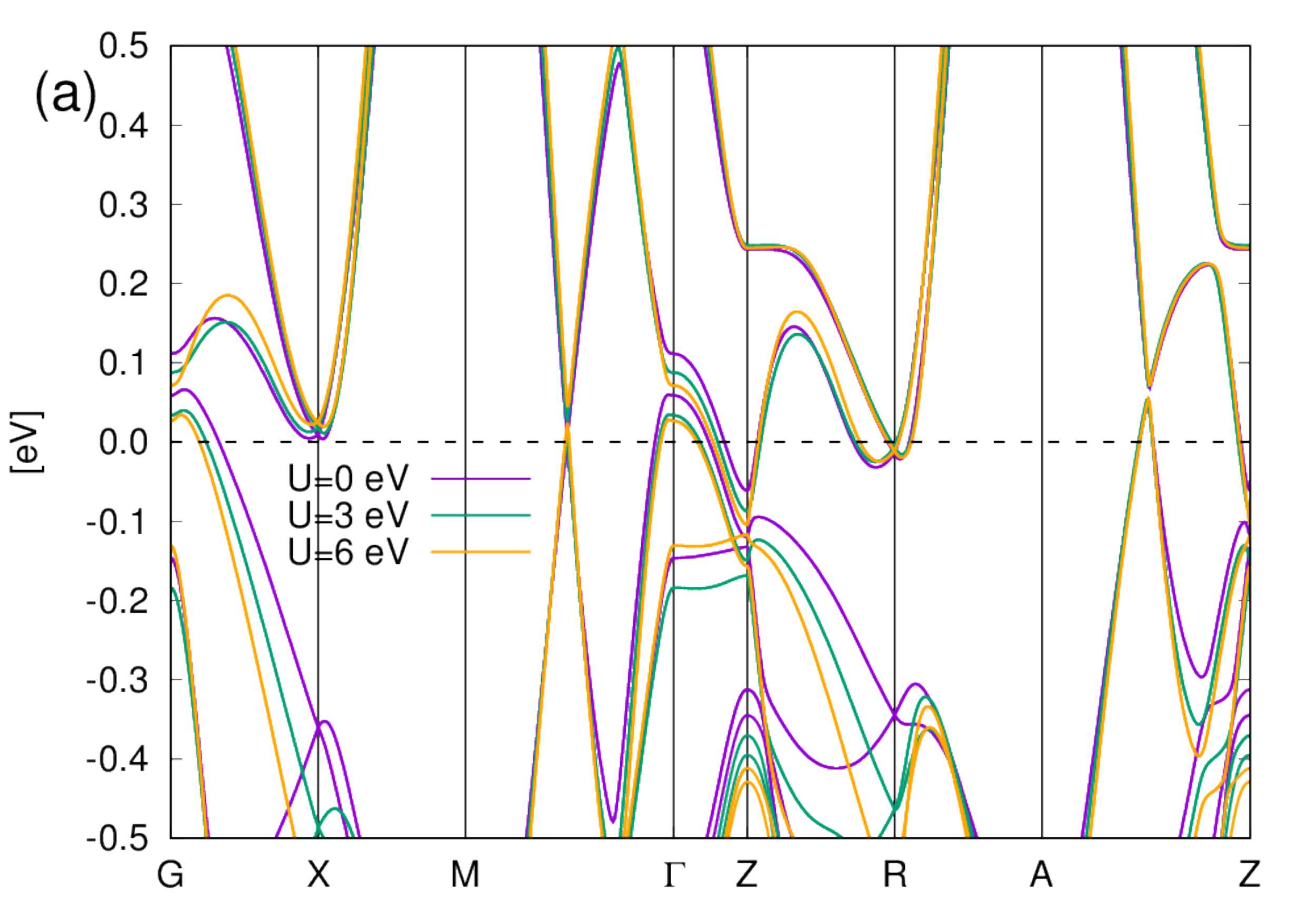}&
 	\includegraphics[width=0.48\columnwidth]{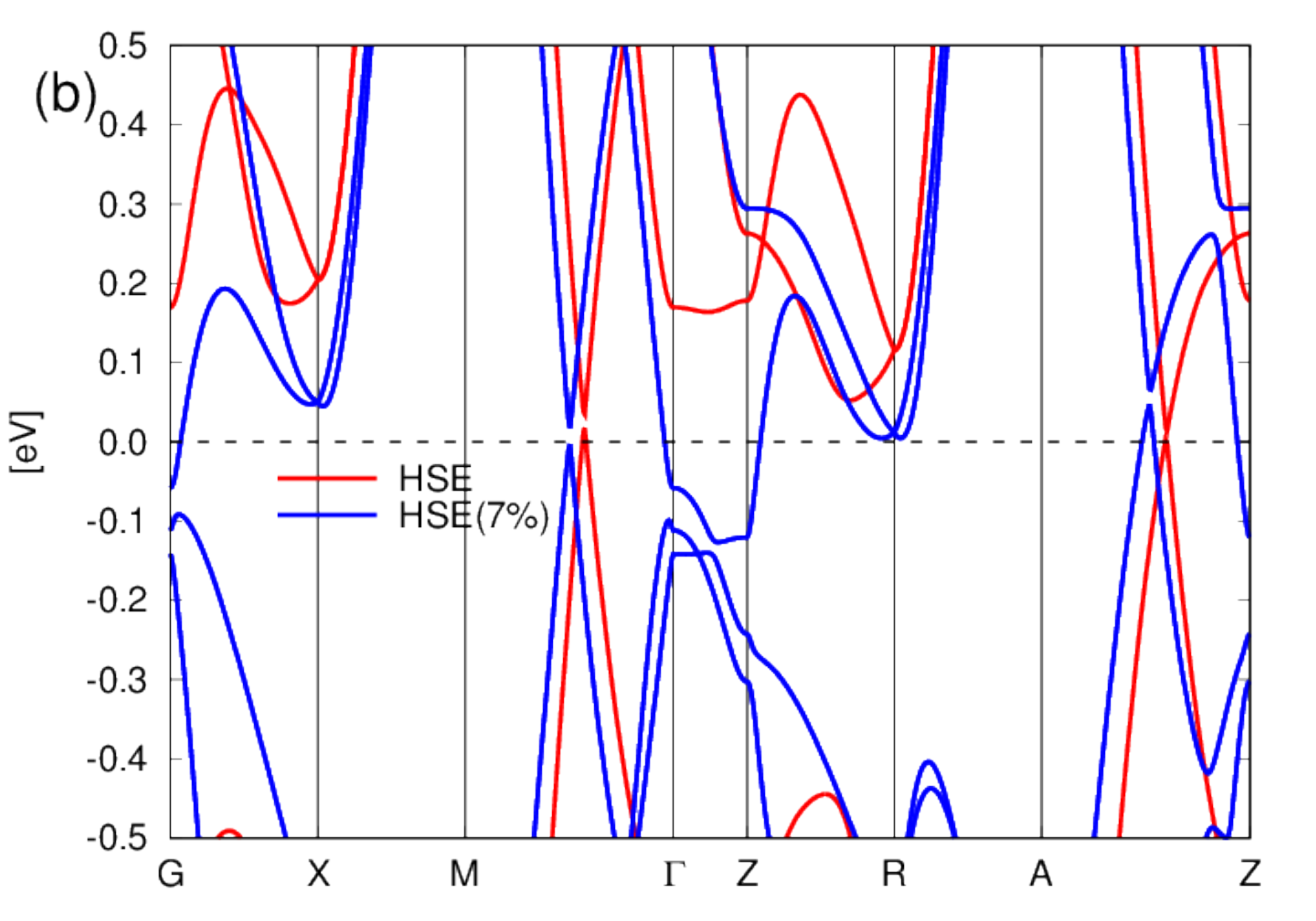}\\
\end{tabular}
\caption{(color online). Panel (a): Band structure obtained with the GGA and GGA+U functionals. The $U$ values are reported in the key. The main difference between the functionals is in the band dispersion along the $\Gamma$-Z path. The larger the $U$, the lower the position of the bands that cross the Fermi level. Panel (b): Band structure obtained with the HSE functional with two fractions of exact exchange. The regular $25 \%$ of exchange is used in ``HSE'', while only the $7 \%$ is used in ``HSE(7\%)''.  Again, the main difference is in the $\Gamma$-Z path. 
\label{bands} 
} 
\end{figure}

\begin{figure}[!h]
\centering 
\begin{tabular}{ccc}  
	\includegraphics[width=0.33\columnwidth]{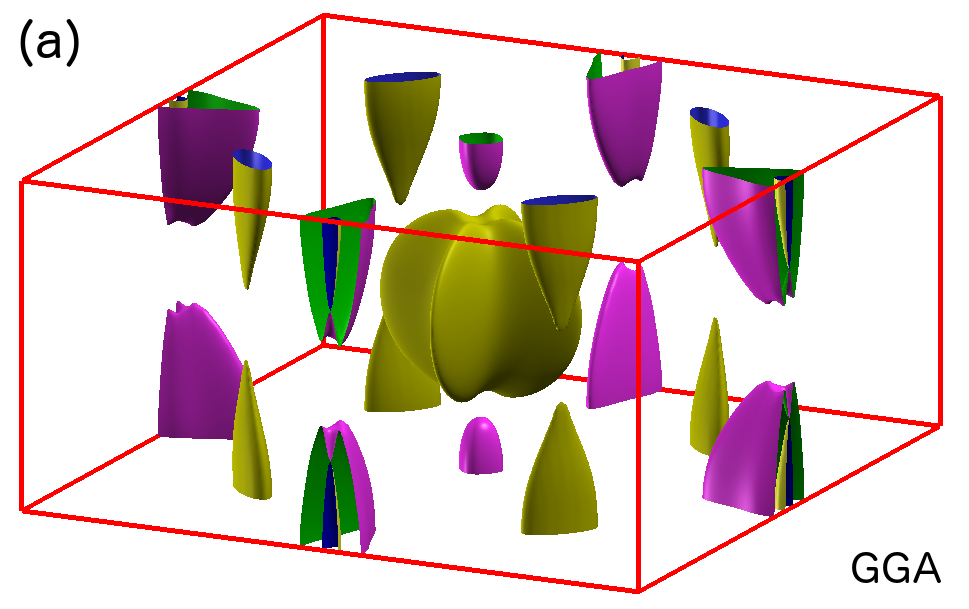}&
 	\includegraphics[width=0.33\columnwidth]{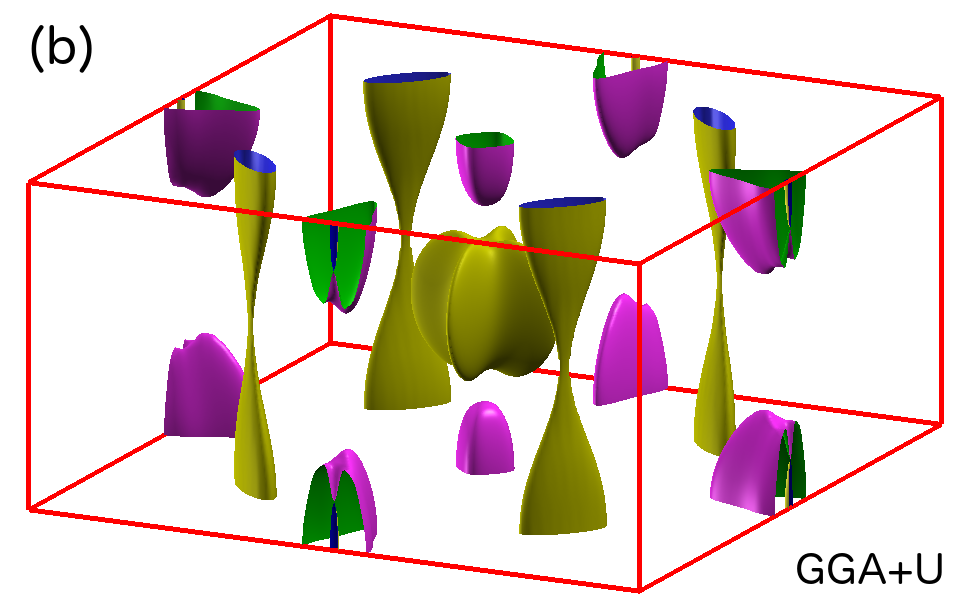} &
    \includegraphics[width=0.33\columnwidth]{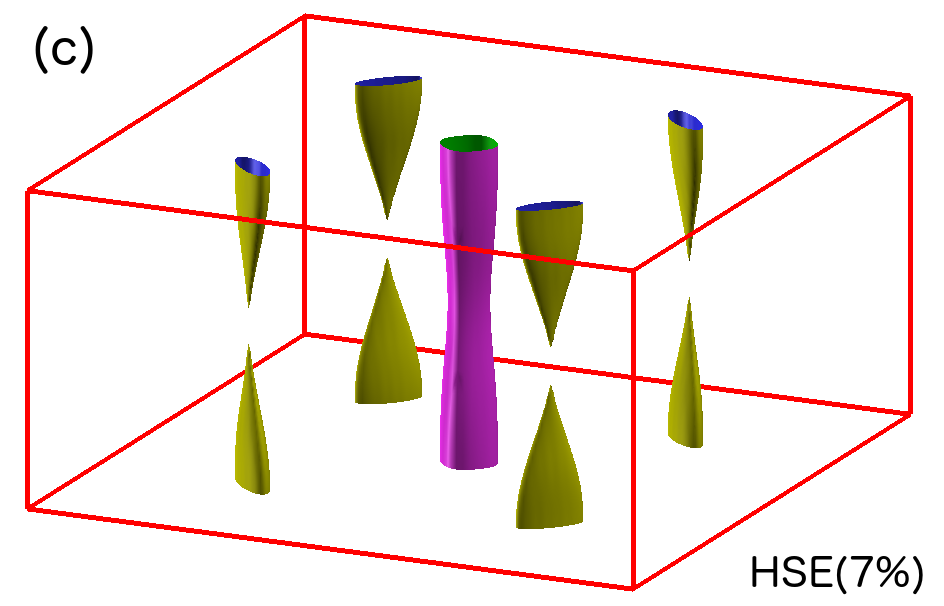}\\
\end{tabular}
\caption{(color online). Fermi surfaces of BaNiS$_2$ from \textit{ab initio} calculations with GGA (panel (a)), GGA+U(=3eV) (panel (b)), and modified HSE with 7\% of exact exchange (panel (c)). The red lines are the frontiers of the first BZ, drawn in Fig.~\ref{fig1}(b) together with the high-symmetry points. The yellow (purple) surfaces indicate Fermi pockets of hole (electron) character.
\label{FS}
} 
\end{figure}

\begin{figure}[!h]
\includegraphics[width=0.75\columnwidth]{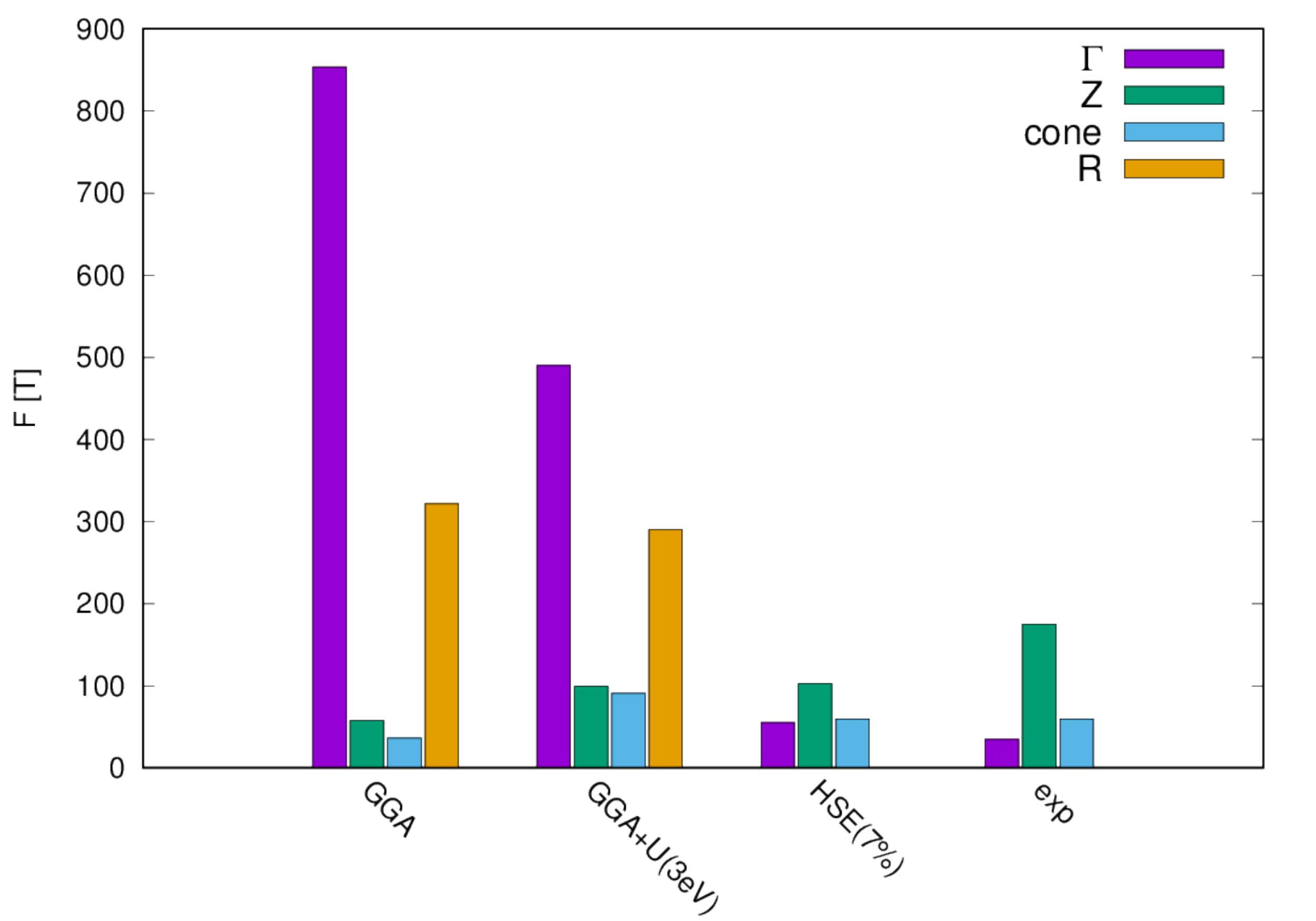}
\caption{(color online). Computed QOs frequencies for $\theta=0$ by employing the same density functionals that give the Fermi surfaces plotted in Fig.~\ref{FS}. Those frequencies correspond to the extremal cross-sections containing the $k$-points reported in the caption. For the modified HSE with $7\%$ of exact exchange (``HSE(7\%)''), there is no pocket around R, at variance with the GGA and GGA+U cases. 
\label{deHaas_ab_initio}} 
\end{figure}

The band structures are reported in Fig.~\ref{bands}, while the corresponding Fermi surfaces are plotted in Fig.~\ref{FS}. Four cones are present in the BZ, whose axes are located at the proximity of the $(-1/4,-1/4,z)$, $(-1/4,1/4,z)$, $(1/4,-1/4,z)$, and $(1/4,1/4,z)$ directions. At $k_z=0$, the Fermi level crosses the cones at their apex. The conical energy dispersion drifts as a function of $k_z$, being lifted up when we move away from $k_z=0$. The dispersion along $k_z$ is caused by the necessity of charge-carrier compensation. This leads to 4 Fermi surfaces that are also conical, instead of being Dirac lines, with vertices lying on the $k_z=0$ plane. These four conical surfaces are a very robust feature of the system, present in \emph{all} DFT calculations performed here. Therefore, they are insensitive to the correlation level, and to the quality of the density functional used. Nevertheless, their related QOs signal, evaluated from the area of the extremal cross-section at $k_z=\pi/c$, varies from functional to functional, as reported in Fig.~\ref{deHaas_ab_initio}. However, their $F$ fluctuation is much milder than for any other Fermi pocket, and it is in a quite good agreement with the experiment for all functionals. Thus, from the \emph{ab initio} values, we can safely attribute the $\beta$ orbit to these 4 cones.
The linear dispersion close to the cone apex is not captured by our experimental results, because it would require the angle dependence of both frequencies and effective mass, as stated in Ref.~\onlinecite{sut11}. However, it is reassuring that our conclusions drawn in Sec. \ref{QOs-results} regarding the carrier concentration remain valid within this picture since, according to the DFT calculations, the volume of the conic part is small. The highly mobile holes observed in magnetotransport measurements are assigned to this conic portion.

In contrast to the robustness of the conical Fermi surfaces, the most fragile Fermi structure is the one along the $\Gamma$-Z direction. As one can evince from Fig.~\ref{bands}(a), the position of the valence and conduction bands, that cross the Fermi level in $\Gamma$-Z, strongly depends on the on-site repulsion strength included in the GGA+U functional. These bands are shifted down by about 50 meV by the addition of a Hubbard term $U = 3$ eV. This shift increases upon increasing the value of $U$. As a result, the hole pocket centered at $\Gamma$ shrinks while the electron pocket centered at Z grows (see Figs.~\ref{FS}(a) and \ref{FS}(b)). Anyway, the corresponding $F$ values reported in Fig.~\ref{deHaas_ab_initio} for the hole pockets at $\Gamma$ are far too large if compared with the experimental findings. Neither GGA nor GGA+U agree with experimental data. Instead, the use of properly tuned hybrid functionals corrects for this pitfall. While regular HSE opens a fictitious gap in the $\Gamma$-Z region, by reducing the exact exchange weight down to 7\% the crossing bands acquire an electron character all the way from $\Gamma$ to Z (Fig.~\ref{bands}(b)). This is reflected in the Fermi surface by the disappearance of the hole pocket at $\Gamma$, as seen in Fig.~\ref{FS}(c). In the modified HSE (``HSE(7\%)''), the electron pocket becomes a 2D constricted tube that stretches throughout the whole BZ along $k_z$. The QOs generated by this structure at $\theta=0$ are given by the extremal cross-sections at $\Gamma$ (minimum) and Z (maximum). Their values are plotted as histograms in Fig.~\ref{deHaas_ab_initio}. They nicely correspond to the experimental values of the $\alpha$ and $\gamma$ orbits, respectively. Therefore, we can identify these two orbits as belonging to the same $\Gamma$-constricted tube, and having $\Gamma$-Z as rotational axis. This maps all measured dHvA frequencies into identified pockets, based on a Fermi surface picture provided by our modified HSE.

It is worth noting that in the GGA and GGA+U band structure, there is yet another electron pocket, centered at $R$ (Fig.~\ref{bands}(a)), which is Rashba active\cite{Santos-Cottin16}. The Rashba splitting gives rise to two nested parabolas, filled by spinors with opposite chirality. One can see the nested shapes in Figs.~\ref{FS}(a) and \ref{FS}(b). This should yield additional frequencies $F$, the strongest being as large as 300 $T$. This value is not found in the experiment. The improved HSE band structure does not have any pocket at $R$, because the conduction bands are raised just above the Fermi level, without touching it. This is another feature that highlights the reliability of the modified HSE functional. At variance, the GGA and GGA+U functionals show several drawbacks, and they are unable to reproduce neither quantitatively nor qualitatively the experimental frequency values. This tells us about the importance of non-local screened-exchange interactions in this material, as we will argue in the concluding Sec.~\ref{conclusions}.

This theoretical analysis allows us to unambiguously assign the $S_1$ surface to the electron-type quasi-2D tube having $\Gamma$-Z as axis with neck and belly centered at $\Gamma$ ($\alpha$ orbit) and Z ($\gamma$ orbit), respectively. On the other hand, the 4 $S_2$ hole-type pockets, corresponding to $\beta$, are assigned to 4 conical Fermi surfaces, whose vertical axes are approximately $(-1/4,-1/4,z)$, $(-1/4,1/4,z)$, $(1/4,-1/4,z)$, and $(1/4,1/4,z)$, with extremal cross-sections located at $k_z=\pi/c$, and apex at $k_z=0$.

\section{Conclusions}
\label{conclusions}

Shubnikov-de Haas and de Haas-van Alphen oscillations are reported for the first time in the undoped BaNiS$_2$ compensated semimetal. The estimated mean free path of 400 \text{\AA} indicates a low concentration of defects in the measured single crystals. We detected three independent main frequencies corresponding to small cyclotron orbits, covering less than 1\% of the FBZ cross-section area. Angle dependences of these frequencies suggest the existence of two types of pockets, which is confirmed by DFT electronic structure calculations when a modified HSE hybrid functional with 7\% of exact exchange is used.

The first type ($S_1$) is a constricted tube of electron-like character and having the $\Gamma$-Z direction as rotational symmetry axis. By symmetry there is just one pocket of this type. The remaining four hole-like pockets belong to the second type ($S_2$). They are oriented in the $z$ direction, and centered in the proximity of one of the $(-1/4,-1/4,z)$, $(-1/4,1/4,z)$, $(1/4,-1/4,z)$, and $(1/4,1/4,z)$ axes. For this second type, the measured frequency indicates a conventional ellipsoidal shape due to parabolic bands. According to band structure calculations the unprobed part of these pockets has a conic-like dispersion and should result in high-mobility carriers. This could be verified experimentally by the determination of the angle dependence of the effective mass.

At first sight, the present results may be in contradiction with our model of the magnetotransport data including three types of charge carriers~\cite{klein16}: minority holes with a high mobility and majority holes and electrons. However, the latter model is based on the Drude law of parabolic bands, which supposes homogeneous effective masses and scattering times of charge carriers, and is not adapted for linear dispersions characteristic of the present hole pockets. It turns out that the recourse to two discrete and homogeneous populations of charge carriers is sufficient to simulate the properties of the conic-like and ellipsoid-like parts of the $S_2$ pockets, hosting hole carriers with different mobilities, very high for the conic-like part, lower in the ellipsoid sector. 

By carrying out \textit{ab initio} electronic structure calculations, we showed that the use of a modified hybrid functional is necessary to explain the experimental values. This modified functional has a reduced ratio of exact exchange, if compared with the standard HSE functional. We found that 7\% of the exact exchange yields a good quantitative agreement with the QOs results. The agreement is lost, even qualitatively, in the case we use a regular GGA functional. Adding the local Hubbard repulsion to the GGA in the GGA+U framework does not improve the GGA picture. This indicates the importance of non-local screened-exchange interactions in the system. Therefore, in the BaNiS$_2$ semimetal, the main role of the electronic correlation is non-local, rather than Hubbard-like.

The role played by the non-local screened exchange can be relevant to other transition metal compounds, such as the iron selenide. Like BaNiS$_2$, FeSe is a multiband compensated semimetal, although its correlation strength is known to be much larger than the one in BaNiS$_2$~\cite{aichhorn2010,Aud15, Ter14, Wat15}. Nevertheless, there are features that are not explained by a local self energy only\cite{brouet2013}. The present study suggests that the non-local screened-exchange interactions cannot be neglected in $3d$ semimetals, particularly when the screening is poor, as a consequence of $d$-electron correlation and charge-carrier compensation. To model these systems, advanced \emph{ab initio} methods are necessary, which include non-local correlations, such as hybrid functionals or many-body methods with non-local self energy, such as GW, GW+DMFT, and their variants\cite{biermann2014,werner2016}. Combining QOs measurements with various types of first-principles calculations turns out to be very important to identify the main renormaliation mechanisms of the low-energy electronic structure.

\begin{acknowledgments}
This work was supported by the University Pierre and Marie Curie under the "Programme \'emergence". The support of the European Magnetic Field Laboratory (EMFL) is acknowledged. MC acknowledges the GENCI allocation for computer resources under the project number A0010906493 and the PRACE allocation under the project PRA143322.
\end{acknowledgments}

\bibliography{biblio}

\end{document}